\documentclass[english]{emulateapj}

\usepackage{textcomp}
\usepackage{amsmath,amssymb,amsfonts}
\usepackage{hyperref}
\usepackage{txfonts}
\usepackage{subfigure}
\usepackage{natbib}

\begin{document}

 \title{\bf{Post main sequence evolution of icy minor planets: II. water retention and white dwarf pollution around massive progenitor stars}}

\author{Uri Malamud and Hagai B. Perets}
\affil{Department of Physics, Technion, Israel}
\email{uri.mal@tx.technion.ac.il ~~~~ hperets@physics.technion.ac.il}

\newpage

\begin{abstract}
Most studies suggest the pollution of white dwarf (WD) atmospheres arises from accretion of minor planets, but the exact properties of polluting material, and in particular the evidence for water in some cases are not yet understood. Here we study the water retention of small icy bodies in exo-solar planetary systems, as their respective host stars evolve through and off the main sequence and eventually become WDs. We explore for the first time a wide range of star masses and metallicities. We find that the mass of the WD progenitor star is of crucial importance for the retention of water, while its metallicity is relatively unimportant. We predict that minor planets around lower-mass WD progenitors would retain more water in general, and would do so at closer distances from the WD, compared with high-mass progenitors. The dependence of water retention on progenitor mass and other parameters has direct implications for the origin of observed WD pollution, and we discuss how our results and predictions might be tested in the future as more observations of WDs with long cooling-ages become available.
\end{abstract}


\section{Introduction}\label{S:Intro}
Despite the typically short sinking time scale of elements heavier than helium in the atmospheres of WDs \citep{Koester-2009}, between 25\% to 50\% of all WDs \citep{ZuckermanEtAl-2003,ZuckermanEtAl-2010,KoesterEtAl-2014} are found to be polluted with heavy elements. This observation is suggestive of their ongoing accretion of planetary material \citep{DebesSigurdsson-2002,Jura-2003,KilicEtAl-2006,Jura-2008}, which is believed to come from perturbed minor planets that survive the main sequence, red giant branch (RGB) and asymptotic giant branch (AGB) stellar evolution phases \citep{DebesSigurdsson-2002,BonsorEtAl-2011,DebesEtAl-2012,FrewenHansen-2014,MustillEtAl-2014,VerasGansicke-2015,Veras-2016}. When crossing the tidal disruption radius of the WD, a circumstellar disk is formed \citep{VerasEtAl-2014a,VerasEtAl-2015}, and eventually accretes onto the WD by various mechanisms \citep{Jura-2008,Rafikov-2011,MetzgerEtAl-2012}.

The spectroscopic analysis of WD atmospheres \citep{WolffEtAl-2002,DufourEtAl-2007,DesharnaisEtAl-2008,KleinEtAl-2010,GansickeEtAl-2012} as well as infrared spectroscopy of the debris disks themselves \citep{ReachEtAl-2005,JuraEtAl-2007,ReachEtAl-2009,JuraEtAl-2009,BergforsEtAl-2014}, are generally consistent with 'dry' compositions, characteristic of inner solar system objects. Only few polluted He-dominated WDs are considered potentially water-rich: GD 61, with 26\% water in mass \citep{FarihiEtAl-2013}, SDSSJ1242, with 38\% water in mass \citep{RaddiEtAl-2015} and WD 1425+540, with 30\% water in mass \citep{XuEtAl-2017}. Also, by looking at en ensemble of 57 nearby He-dominated WDs, \cite{JuraXu-2012} concluded that the summed hydrogen in their atmospheres must have been delivered by very dry bodies, the water mass fraction not exceeding 1\%. Among a subset of oxygen-containing WDs in the above-mentioned ensemble, no more than 5.8\% of this oxygen could have been carried in water \citep{JuraYoung-2014}. Similar conclusions have been reached by \cite{PietroGentileFusilloEtAl-2017}, applying the same analysis on a larger and more recent sample of He-dominated SDSS WDs from \cite{KoesterKepler-2015}. Therefore, observations so far would seem to point towards relative scarcity of water in polluted WDs. This remains an important ongoing question.

Since the most likely origins of WD pollution is accretion of minor planets in evolved WD systems, understanding the structure and composition of such bodies throughout the evolution of their host stars up to, and during the WD stage is therefore crucial for understanding and constraining models of WD pollution, and for the interpretation of the observed composition of polluted WD atmospheres. \cite{JuraXu-2010} have made the first pioneering model for the evolution of such planetesimals. In a predecessor paper \citep{MalamudPerets-2016} (or MP16 henceforward), we followed-up on this work and improved on it by utilizing a new and significantly more sophisticated evolution code for icy minor planets, that considered new parameters which were previously unaccounted for. We found that water can survive in a variety of circumstances and locations around evolved sun-like stars, with water retention generally increasing with orbital distance, and discussed the implications for WD pollution. 

Given the scope, level of detail, high number of free model parameters, and the entailed discussion in our previous paper (MP16), the work had to be limited by restricting stellar evolution only to a 1 M$_{\odot}$, sun-like (solar metallicity) star. The retention of water in planetary systems around more massive stars was not investigated. However, observed WDs typically have masses of $\sim$0.6 M$_{\odot}$ arising from higher mass progenitors (sun-like stars evolve into $\sim$ 0.5 M$_{\odot}$ WDs), with a distribution extending up to 1-1.3 M$_{\odot}$. Although most WDs are in the mass range 0.5-0.8 M$_{\odot}$, the number of WDs in the tail is not negligible. In order to complement our previous work, the purpose of this paper is therefore to consider a wider range of progenitor masses and metallicities, investigating the fate of their respective water-bearing minor planets. In what follows we briefly outline in Section \ref{S:Model} the model used in this study. In Section \ref{S:Results} the results of our model are presented, and discussed in Section \ref{S:discussion}.

\section{Model}\label{S:Model}
Our evolution model couples the thermal, physical and chemical evolution of icy minor planets of various sizes. It considers the energy contribution primarily from radiogenic heating, latent heat released/absorped by geochemical reactions and surface insolation. It treats heat transport by conduction and advection, and follows the transitions among three phases of water (crystalline ice, liquid and vapor), and two phases of silicates (hydrous rock and anhydrous rock). The model is exactly identical in all aspects to the one used in our predecessor paper. See MP16 for a full and comprehensive review of the model details (equations, parameters, numerical scheme, etc.) and how it compares to previous, simpler models in the literature. The main difference in this work is that the change in star luminosity and mass as a function of time, used as input in our minor planet evolution code, is calculated for more massive progenitor stars and various metallicities. The change in star luminosity then determines the surface boundary condition for the minor planet, whereas the rate of its orbital expansion (dictated by conservation of angular momentum) is determined by the decrease in star mass (primarily during the AGB phase). In order to generate this input we use the MESA stellar evolution code \citep{PaxtonEtAl-2011}. 

The early work of \cite{JuraXu-2010} on the retention of water in post main sequence planetary systems considers two stellar masses: 1 M$_{\odot}$, as in the more recent work of MP16, and 3 M$_{\odot}$. Both with stellar metallicity. They assume that these progenitor masses account for the majority of observed WDs. This is a very reasonable assumption, although the exact relation between the initial progenitor mass to the final WD mass is still not fully constrained. Estimates are based on theoretical stellar evolution models, semi-empirical evidence from open star clusters and observational constraints from wide double white dwarfs. A comparison between the three methods is given by \cite{AndrewsEtAl-2015}. In theoretical stellar evolution models, this ratio also depends heavily on the choice of certain parameters, particularly the metallicity. A lower metallicity correlates with a shorter stellar evolution time, a higher luminosity and a more massive WD. Other evolution parameters are also important \citep{Marigo-2013}, but to a lesser degree. To alleviate some of these concerns, we consider in this study a wider range of progenitor masses, as well as different metallicities. We wish to explore WD masses that correspond to the full range of the mass distribution. In the previous paper we explored the stellar evolution of a 1 M$_{\odot}$, sun-like progenitor, with solar metallicity, corresponding to a low WD mass of 0.52 M$_{\odot}$. Here we consider progenitor masses of 1, 2, 3, 3.6, 5 and 6.4 M$_{\odot}$, with a metallicity of 0.0143 (or [Fe/H]=0 -- the typically used iron abundance relative to solar) corresponding to the final WD masses of 0.54, 0.59, 0.65, 0.76, 0.89 and 1 M$_{\odot}$ respectively. These WD masses probe primarily the peak of the WD mass distribution, but also the tail, at regular mass intervals. WD masses of 0.5 or less are ignored since their progenitor stars should have a mass just under 1 M$_{\odot}$, hence their stellar evolutions should be very similar to that of our minimal stellar mass.

In addition to spanning the range of WD masses, we also explore for the first time the effect of metalicity. We however limit our investigation to a 2 M$_{\odot}$ progenitor, whose resulting WD mass corresponds approximately to the peak mass in the WD mass distribution. We investigate, in addition to the previous metallicity of 0.0143, a one order of magnitude reduction, as well as two orders of magnitude reduction (i.e., z=0.0143, 0.00143,0.000143, or [Fe/H]=0,-1,-2). These metallicities correspond to the final WD masses of 0.59, 0.61 and 0.66 M$_{\odot}$. We extract the luminosity and mass as a function of time for our various choices of initial progenitor masses and metalicities from an available compilation of MESA evolutionary tracks by \cite{ChoiEtAl-2016}.

The outcome of the main-sequence, RGB and AGB stellar evolution phases in terms of water retention, depends on five different parameters. As suggested above, the first parameter is the progenitor's mass. More massive progenitors correlate with a more luminous stellar evolution, albeit a shorter lifetime and also a higher initial (progenitor) to final (WD) mass ratio. The former affects water retention negatively, while the latter two have a positive effect. We also investigate four additional variables related to the minor planets themselves: size (radius), orbital distance, formation time and initial rock/ice mass ratio. To comply with our previous work, we consider exactly the same parameter space, apart from the initial orbital distances, as outlined below. 

(1) Object radius - we consider the following radii: 1, 5, 25, 50 and 100 km. This covers the entire size spectrum from small comets to moonlet sized objects.

\begin{figure*}
	\begin{center}
		\subfigure[1 M$_{\odot}$ progenitor - 0.54 M$_{\odot}$ WD - 11.4 Gyr evolution] {\label{fig:1sm}\includegraphics[scale=0.5]{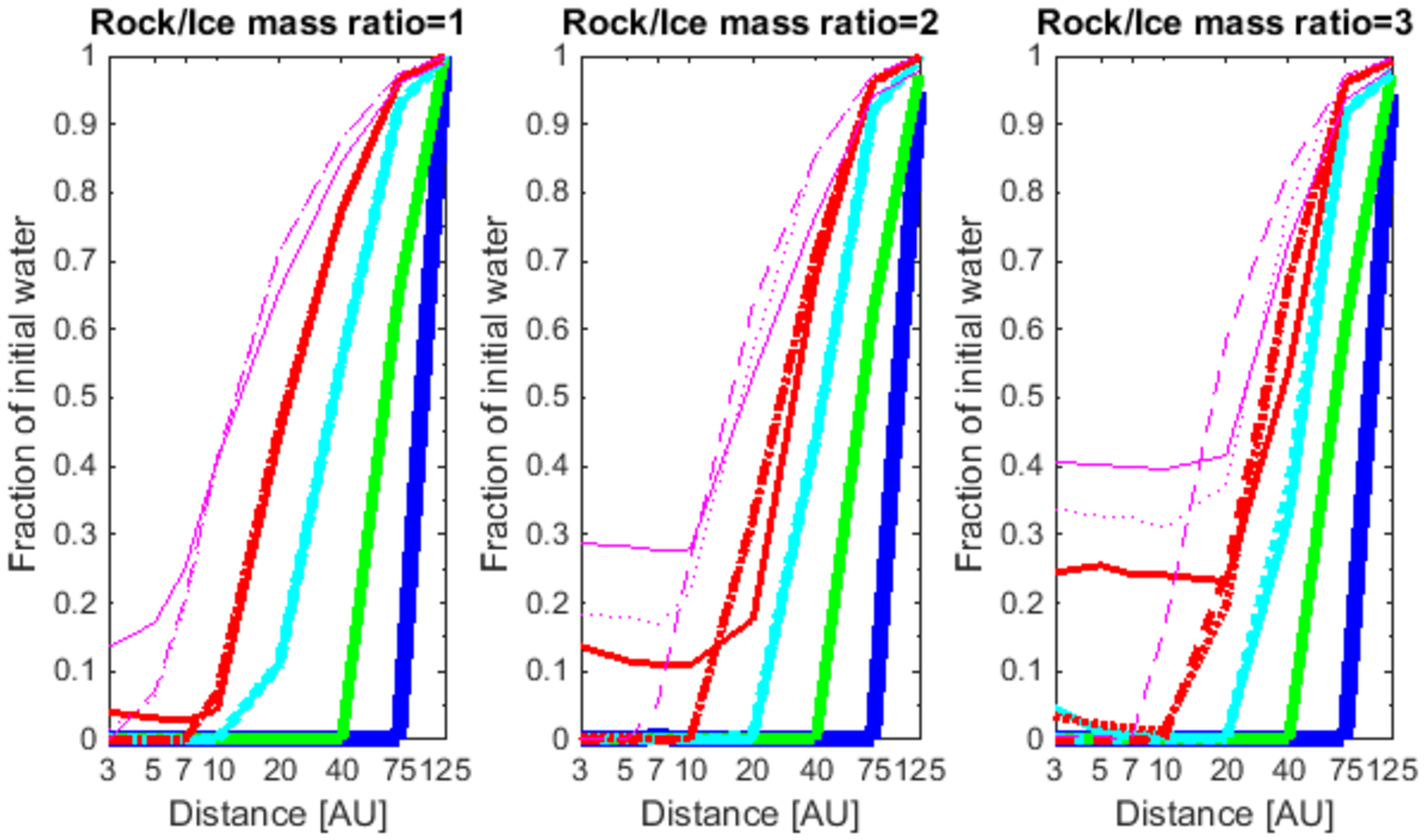}}
		\subfigure[2 M$_{\odot}$ progenitor - 0.59 M$_{\odot}$ WD - 1.34 Gyr evolution] {\label{fig:2sm}\includegraphics[scale=0.5]{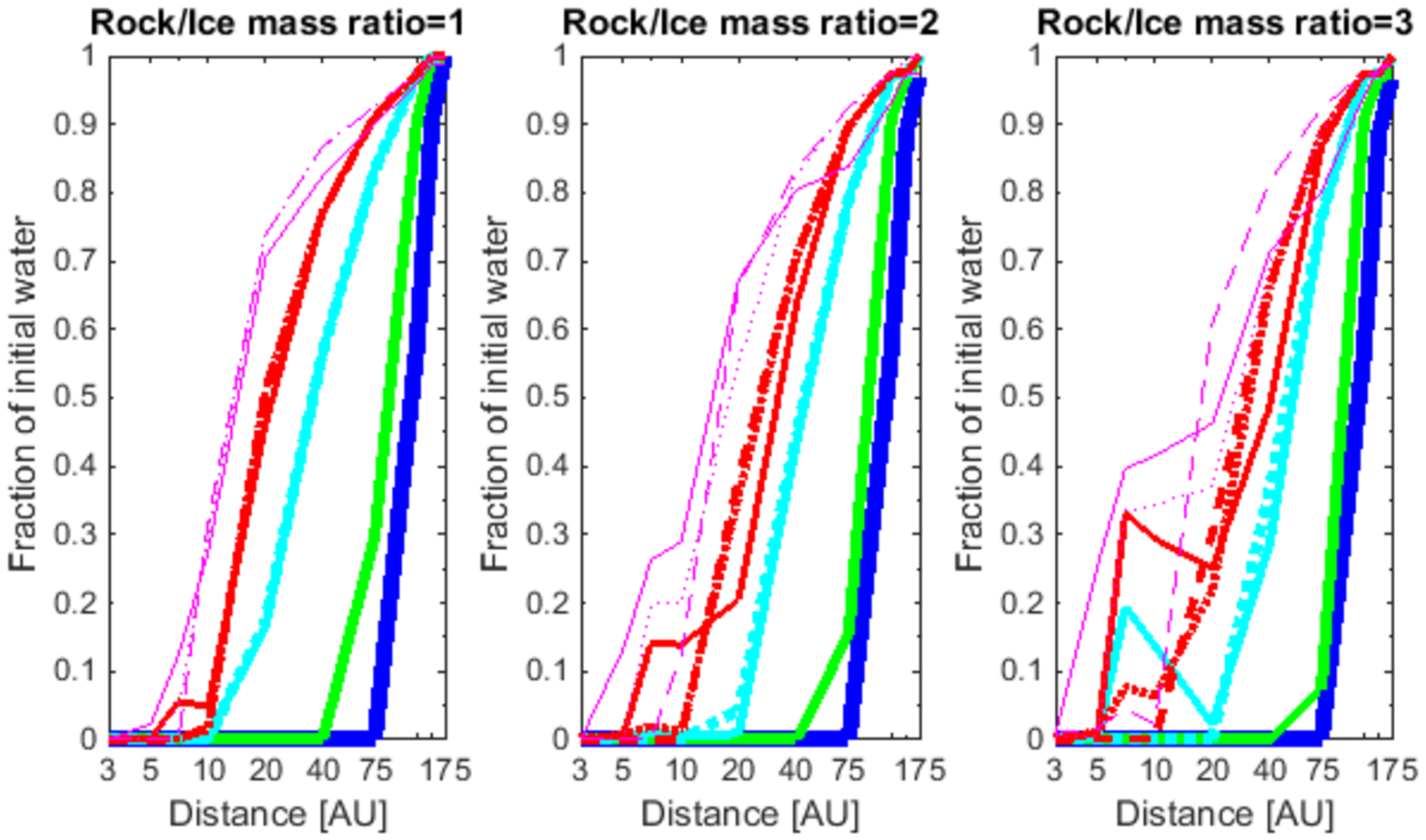}}
		\subfigure[3 M$_{\odot}$ progenitor - 0.65 M$_{\odot}$ WD - 472 Myr evolution] {\label{fig:3sm}\includegraphics[scale=0.5]{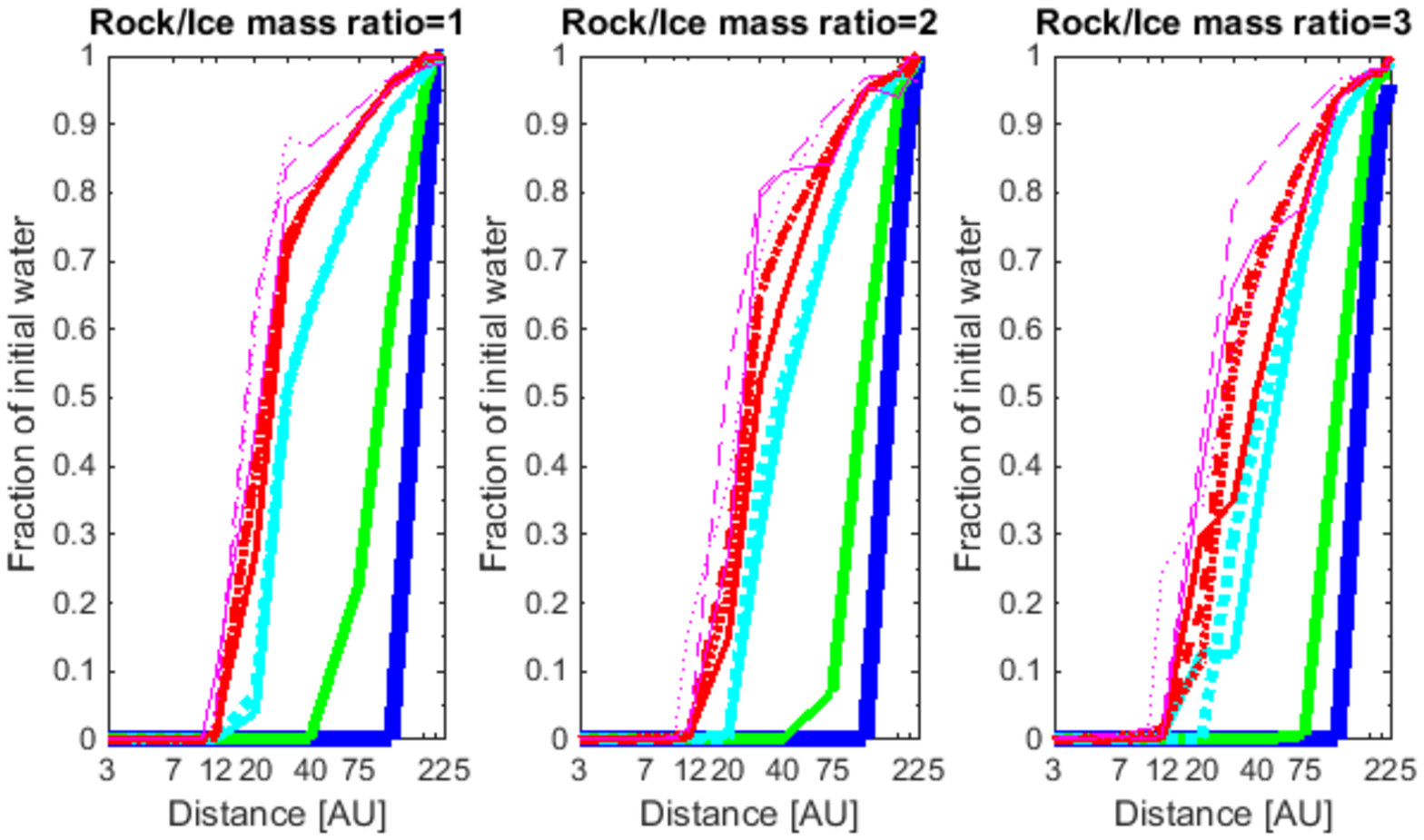}}
		\subfigure[3.6 M$_{\odot}$ progenitor - 0.76 M$_{\odot}$ WD - 278 Myr evolution] {\label{fig:36sm}\includegraphics[scale=0.5]{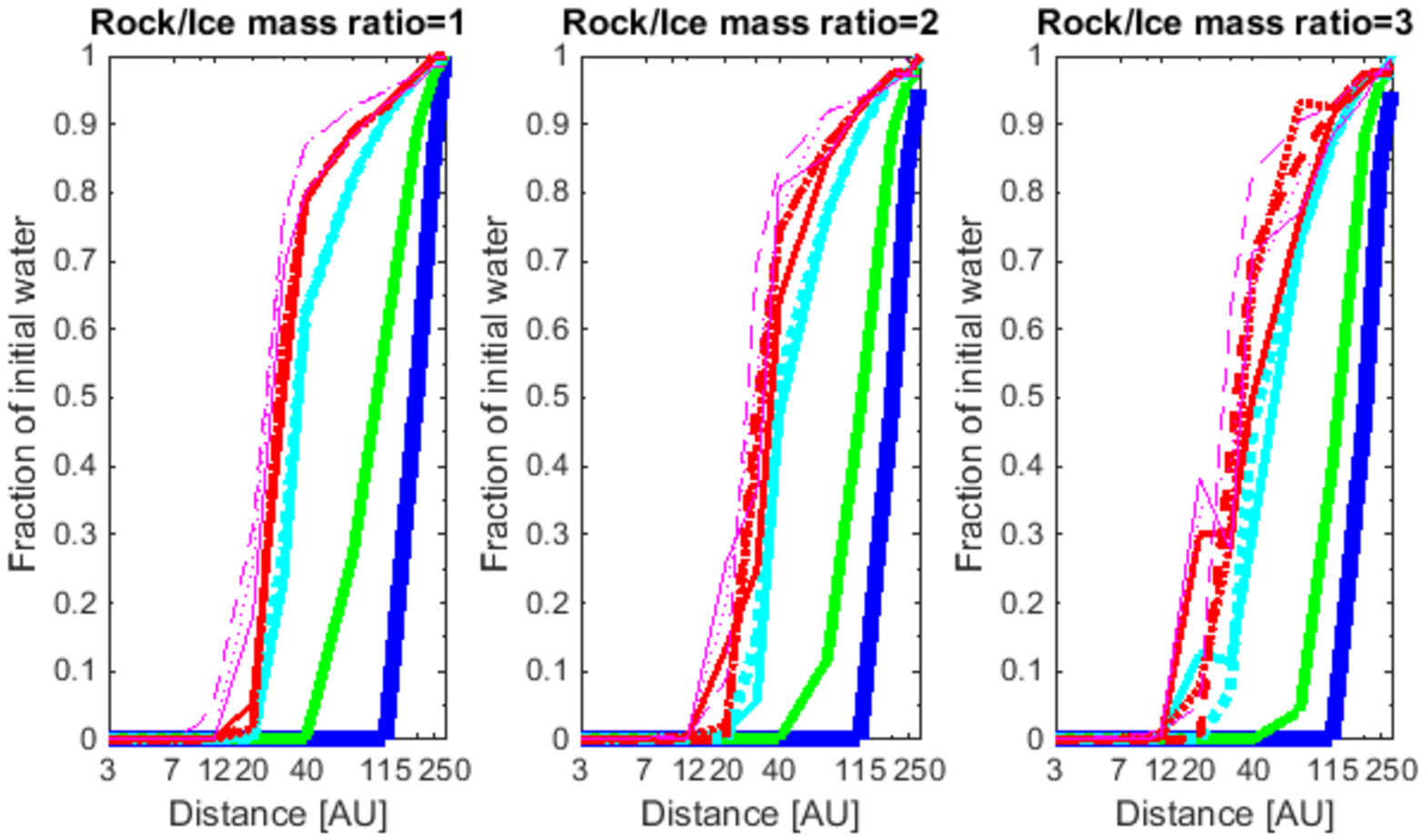}}
		\subfigure[5 M$_{\odot}$ progenitor - 0.89 M$_{\odot}$ WD - 117 Myr evolution] {\label{fig:5sm}\includegraphics[scale=0.5]{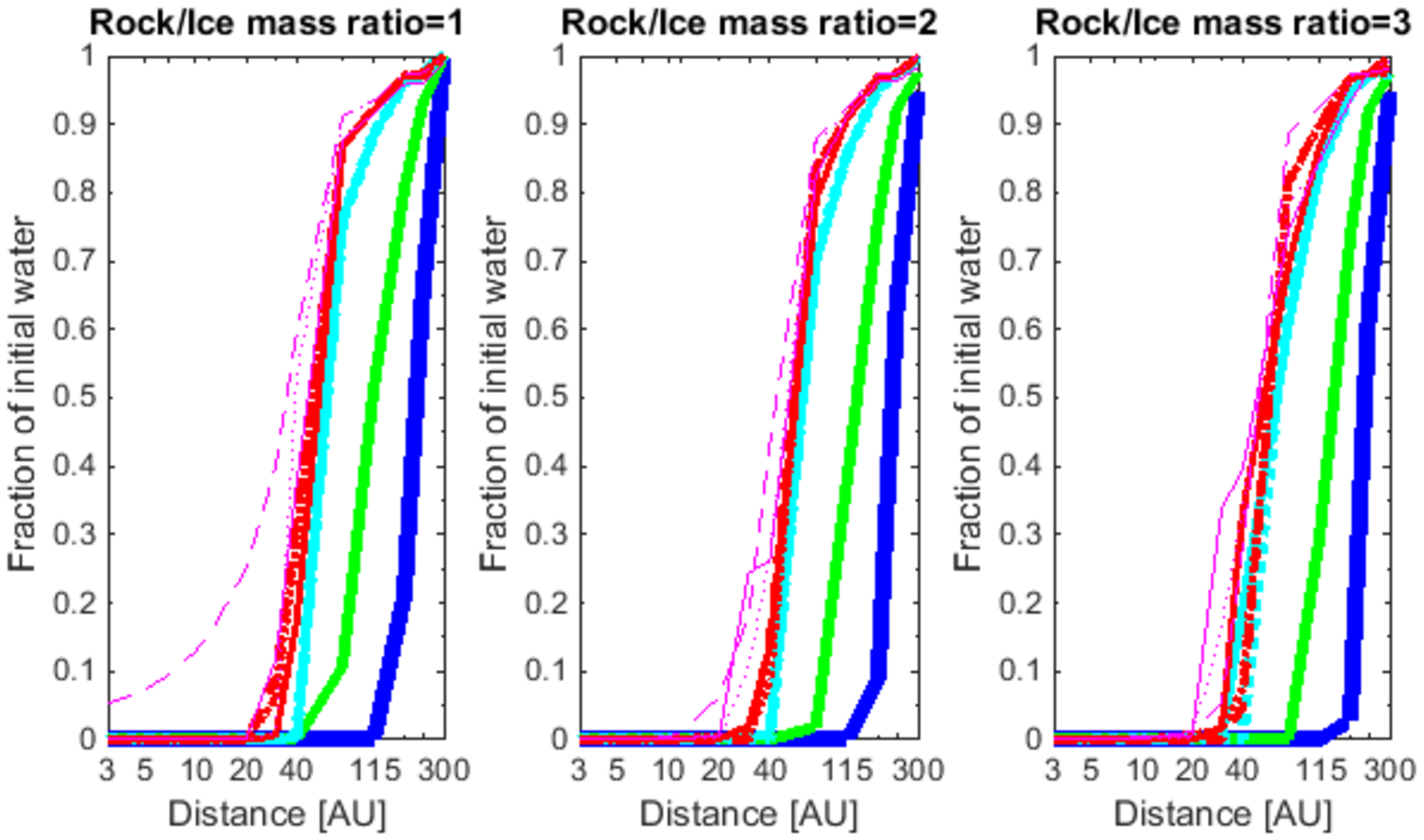}}
		\subfigure[6.4 M$_{\odot}$ progenitor - 1 M$_{\odot}$ WD - 65 Myr evolution] {\label{fig:64sm}\includegraphics[scale=0.5]{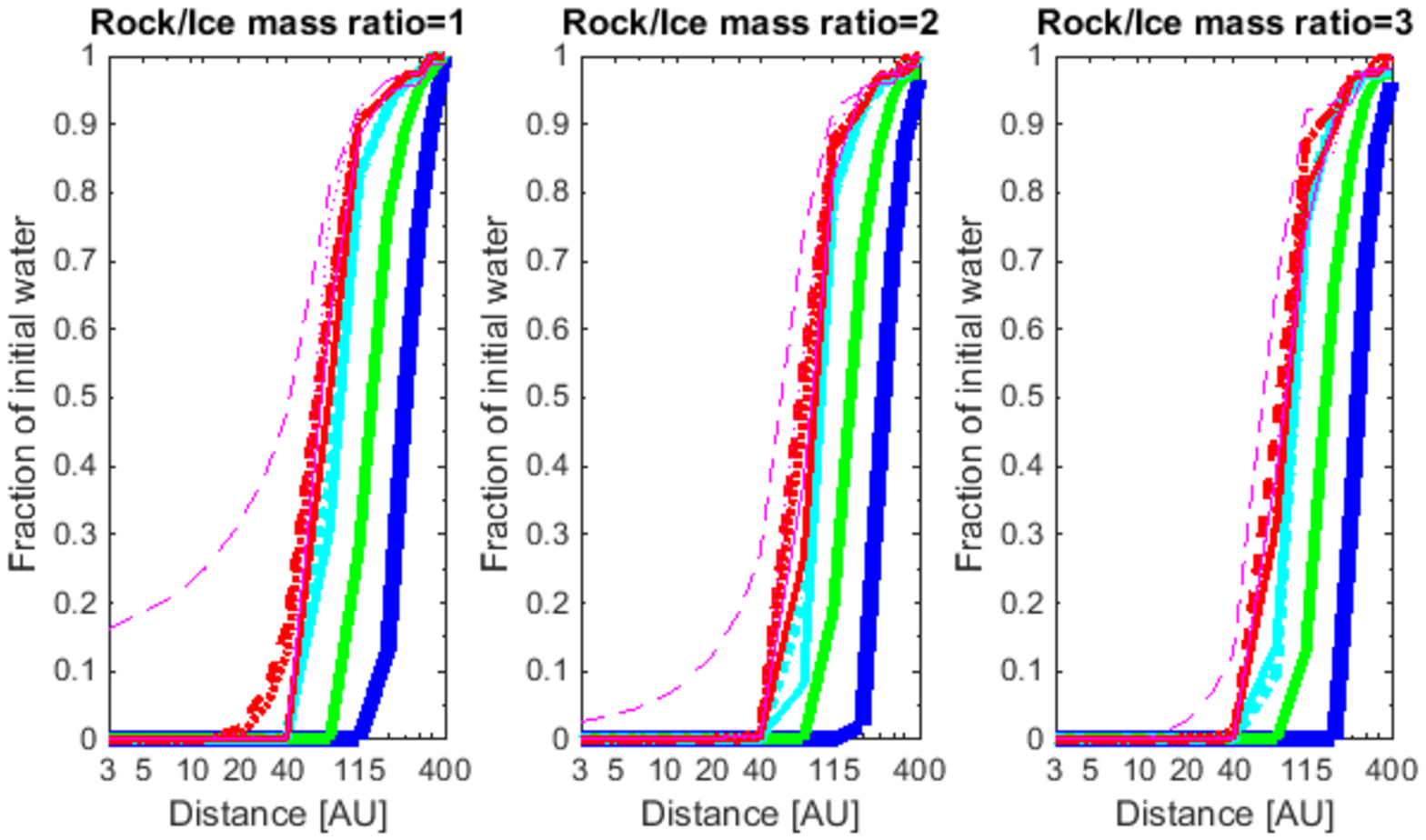}}
		\subfigure{\includegraphics[scale=0.5]{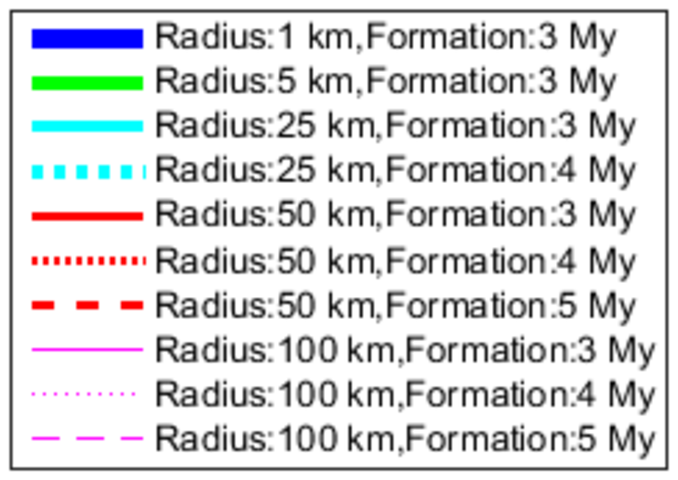}}
	\end{center}
	\caption{Total fraction of water (ice + water in hydrated silicates) remaining after the main sequence, RGB and AGB stellar evolution phases, for different progenitor masses with solar iron abundance [Fe/H]=0. The retention of water is shown as a function of the minor planet's initial orbital distance, composition, radius and formation time.}
	\label{fig:Masses}
\end{figure*}

(2) Orbital distance - We consider a range of possible initial orbital distances (note that with stellar mass loss the orbit undergoes expansion as the minor planet conserves its angular momentum). The minimal initial orbital distance is 3 AU. Below approximately 3 AU, a massive planet (and by extension also its moons if they exist) runs the risk of being engulfed or otherwise tidally affected by the expanding envelope of the post-main sequence RGB \citep{KunitomoEtAl-2011,VillaverEtAl-2014} or AGB \citep{MustillVillaver-2012} star. Conversely, small, asteroid-sized minor planets (of order $\sim$10 km) may inspiral into the star due to dynamical wind drag \citep{Jura-2008}. This might also be an important effect up to $\sim$3 AU. Another consideration for icy minor planets is the location of the snowline \citep{KennedyKenyon-2008}, although minor planets could potentially also migrate inward from their initial birthplace. Overall, a minimum of $\sim$3 AU seems appropriate. We consider the upper bound to be determined by the distance at which full water retention is ensured even for the smallest (radius=1 km) objects. Given a range of progenitor masses however, this upper bound must change, as seen in Fig. \ref{fig:Masses}. It tends to increase with progenitor star mass, and therefore the number of grid points increases accordingly to cover a wider distance range.

(3) Formation time - the formation time of a minor planet is defined as the time it takes a minor planet to fully form, after the birth of its host star. Since here we only consider first-generation minor planets, this time is usually on the order of $\sim 10^0-10^1$ Myr. The formation time determines the initial abundance of short-lived radionuclides, and thus the peak temperatures (hence, internal structure) attained during its early thermal evolution. Although it is clear that the formation time also depends on the orbital distance, the exact relation is unconstrained, which is why we set the formation time as a free parameter. We consider the following formation times: 3, 4 and 5 Myr, complying with our previous work (for our choice of minor planet sizes, shorter or longer formation times were found to be redundant). The initial abundances of radionuclides are assumed to be identical to the canonical values in the solar system, for lack of a better assumption.

(4) Initial rock/ice mass ratio - this ratio initially depends on the location of the object as it forms in the protostar nebula, and like the formation time this parameter is unconstrained (see MP16 for various estimations). We consider three initial rock/ice mass ratios to allow for various possibilities: 1, 2 and 3 (that is, a rock mass fraction of 50\%, 67\% and 75\% respectively).

The total number of models for a single stellar evolution is thus determined by the number of variable parameters (5 x (7-15) x 3 x 3) amounting to several hundreds. However, objects with 1 and 5 km radii are sufficiently small that formation times of 4 and 5 Myr are indistinguishable from a 3 Myr formation time. Objects with a 25 km radius are sufficiently small that a formation time of 5 Myr is indistinguishable from a 4 Myr formation time. These models are thus not contributing any new information and are omitted. Overall, considering all the relevant parameter combinations, we have more than two and a half thousand production runs in total. These models were calculated using a cluster computer. The typical run time of each model was on the order of several hours on a single 2.60GHz, Intel CPU. Except for the variable parameters mentioned above, all other model parameters are equal, and identical to the parameters used in the previous study (MP16), listed in their Table 2.

\section{Results}\label{S:Results}
\begin{figure}
\begin{center}
    \label{fig:2-7-3-3}\includegraphics[scale=1]{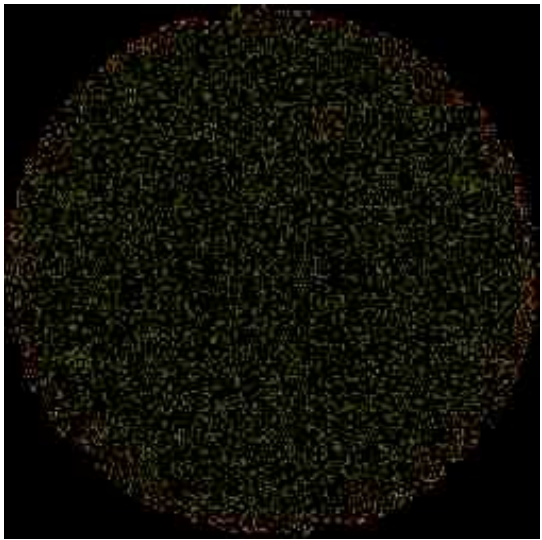}
	\caption{Animated figure (duration - 20 s) featuring the first 10 Myr of evolution (from a total of 1.35 Gyr) of a 100 km radius minor planet around a 2 M$_{\odot}$ progenitor at 7 AU, with a 3 Myr formation time and 75\% initial rock fraction. Colour interpretation: {\it black} (pores); {\it white} (water ice); {\it blue} (liquid water); {\it brown} (anhydrous rock); and {\it olive} (hydrated rock). The animation shows the differentiation of an initially homogeneous body into a hydrous rocky core underlying a thin ice-enriched crust, from which the ice subsequently sublimates.}
\end{center}
\end{figure}

\begin{figure}
	\begin{center}
	{\label{fig:3-12-4-3}\includegraphics[scale=1]{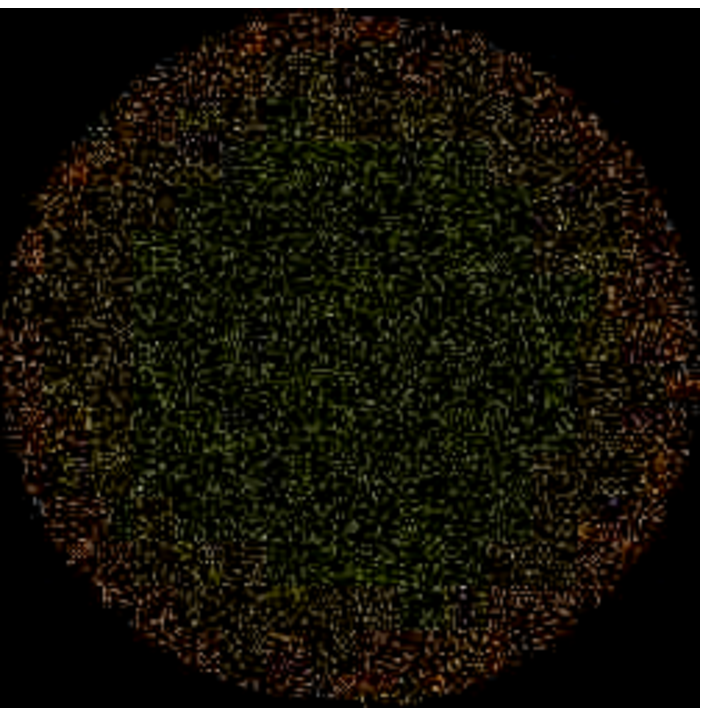}}
	\caption{Animated figure (duration - 22 s) featuring the first 1.7 Myr of evolution (from a total of 470 Myr) of a 100 km radius minor planet around a 3 M$_{\odot}$ progenitor at 12 AU, with a 4 Myr formation time and 75\% initial rock fraction. Colour interpretation: {\it black} (pores); {\it white} (water ice); {\it blue} (liquid water); {\it brown} (anhydrous rock); and {\it olive} (hydrated rock). The animation shows the differentiation of an initially homogeneous body into a smaller hydrous rocky core underlying a thicker ice-enriched crust (both compared to Fig. \ref{fig:2-7-3-3}). Here the inside-out migration of water and outside-in sublimation of water ice from the mantle, occur on similar time scales.}
\end{center}
\end{figure}

\begin{figure}
	\begin{center}
	{\label{fig:6.4-5-5-2}\includegraphics[scale=1]{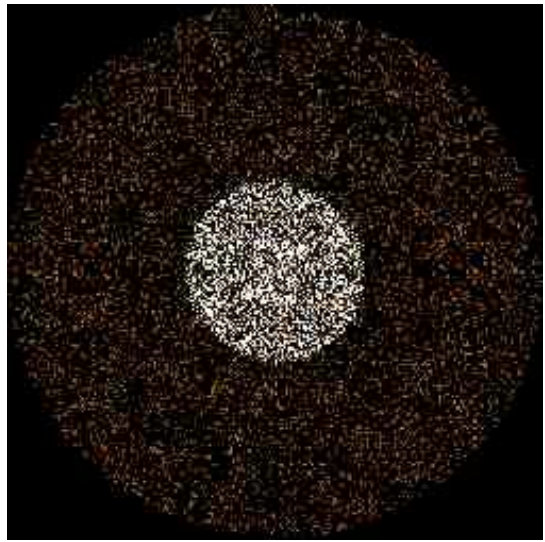}}
	\caption{Animated figure (duration - 21 s) featuring the full 64 Myr evolution of a 100 km radius minor planet around a 6.4 M$_{\odot}$ progenitor at 5 AU, with a 5 Myr formation time and 67\% initial rock fraction. Colour interpretation: {\it black} (pores); {\it white} (water ice); {\it blue} (liquid water); {\it brown} (anhydrous rock); and {\it olive} (hydrated rock). The animation shows the gradual ablation of ice from the surface inwards. This body never differentiates, and its host star's evolution is too short to remove all its internal ice.}
\end{center}
\end{figure}

\begin{figure}
	\begin{center}
		\subfigure[Iron abundance {[Fe/H]}=-1 - 0.61 M$_{\odot}$ WD - 916 Myr evolution]
		{\label{fig:2lm}\includegraphics[scale=0.5]{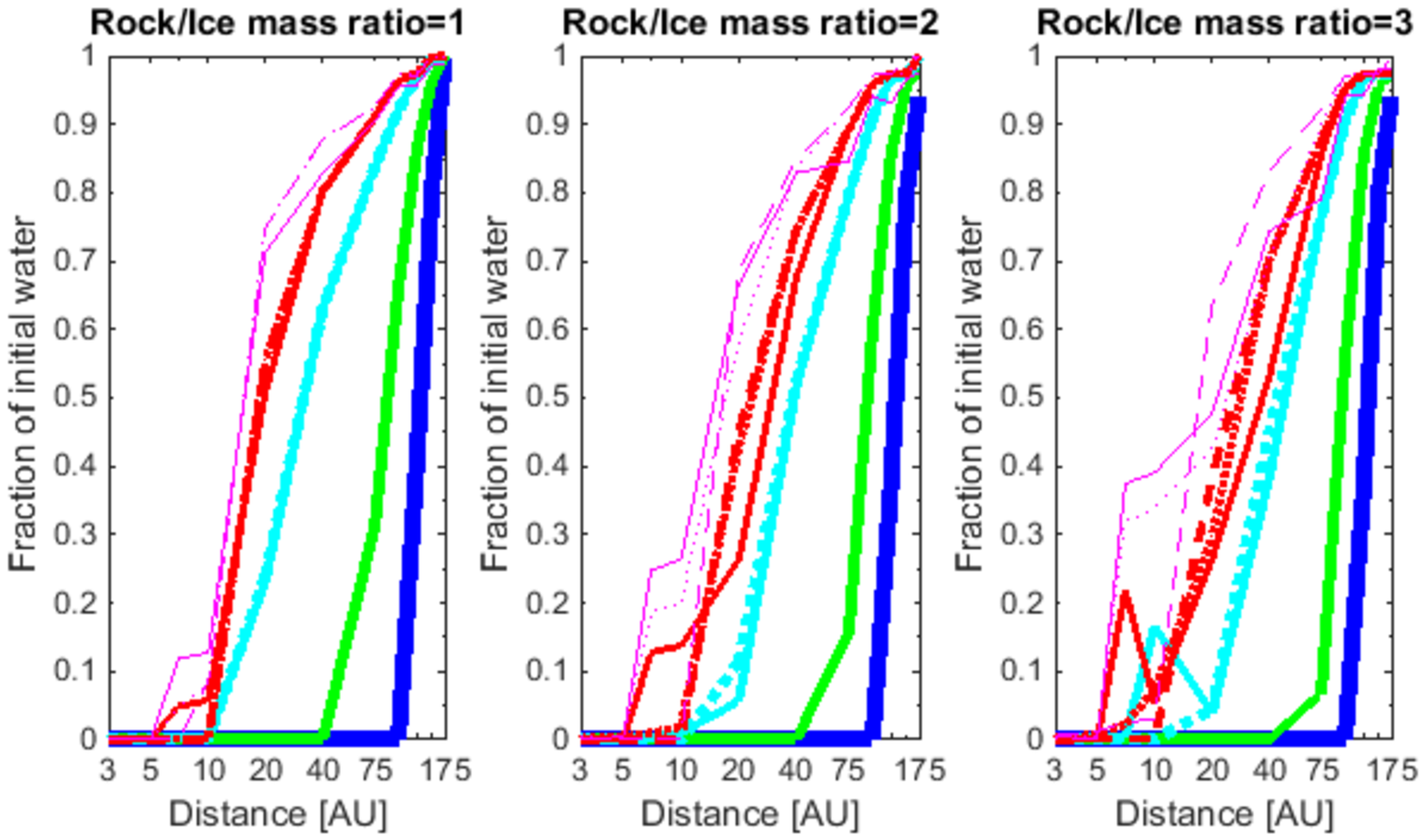}}
		\subfigure[Iron abundance {[Fe/H]}=-2 - 0.66 M$_{\odot}$ WD - 767 Myr evolution]
		{\label{fig:2vlm}\includegraphics[scale=0.5]{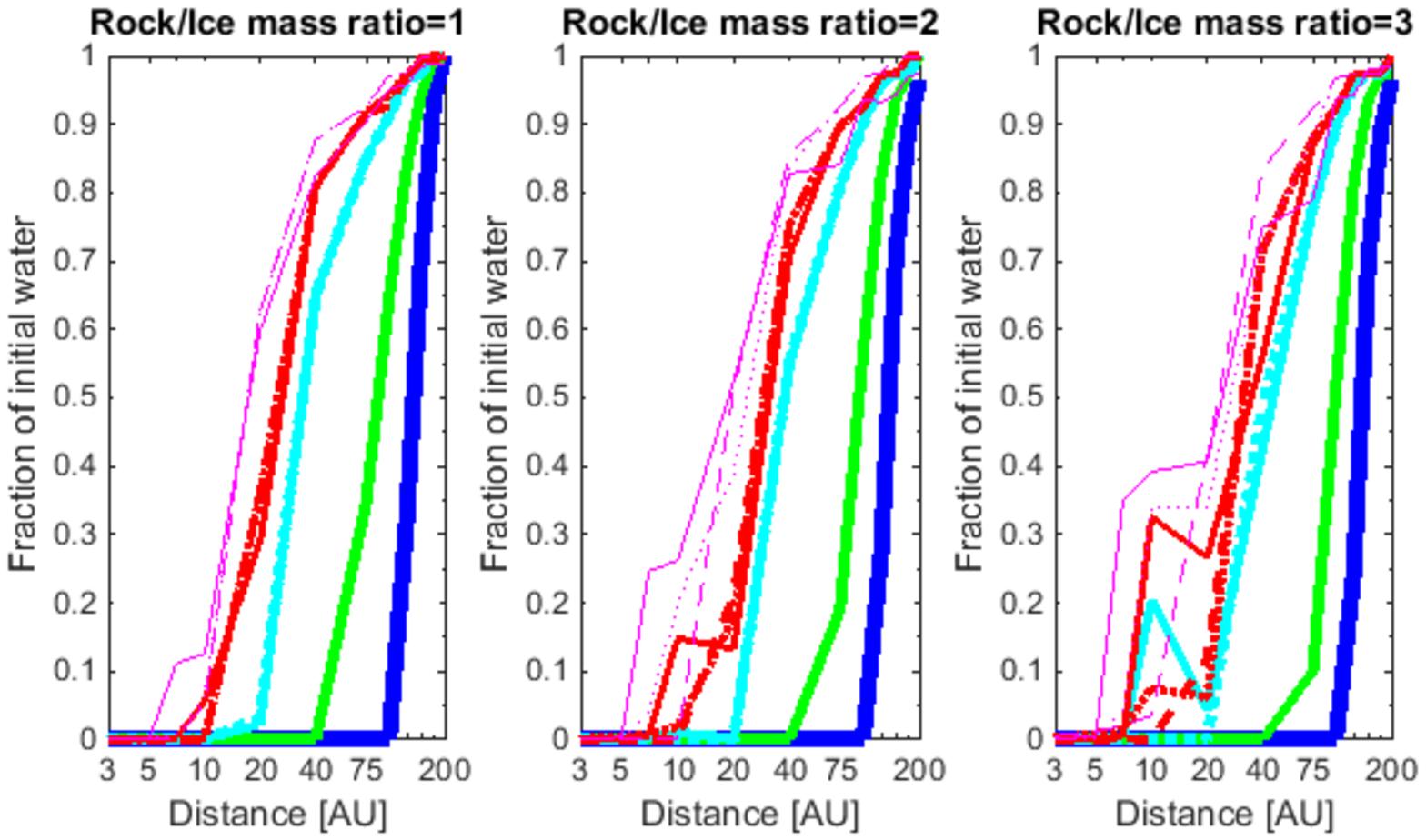}}
		\subfigure{\includegraphics[scale=0.5]{legend.eps}}
	\end{center}
	\caption{Total fraction of water (ice + water in hydrated silicates) remaining after the main sequence, RGB and AGB stellar evolution phases, for a 2 M$_{\odot}$ progenitor with reduced iron abundances. The retention of water is shown as a function of the minor planet's initial orbital distance, composition, radius and formation time.}
	\label{fig:Metallicities}
\end{figure}

In this section we discuss the bulk amount of water surviving in the planetary system, as a function of our free model parameters. Figs. \ref{fig:Masses}-\ref{fig:Metallicities} show the final fraction of water, based on the end states of the production runs discussed in Section \ref{S:Model} (i.e., when the star reaches the WD stage). We present the \emph{total} fraction of retained water, defined as water ice + water in hydrated silicates, which ultimately contributes hydrogen and oxygen when accreting onto polluted WD atmospheres. Each panel consists of three subplots, each representing a different choice for the initial composition. Within each subplot there are multiple lines, depicting the final water fraction as a function of the initial orbital distance. Each line is characterized by a specific color and width, as well as a style. The line width decreases with the size of the object, so thin lines represent large objects, and each line style corresponds to a different formation time. 

Fig. \ref{fig:Masses} summarizes the main results from this study. Note that Panel \ref{fig:1sm} (upper-left) is nearly identical to the equivalent panel in our previous study (Fig. 4(a) from MP16, having the same stellar mass, however a slightly higher metallicity compared to [Fe/H]=0). All the other panels however indicate a notable change in water retention as the progenitor mass increases. 

Another important model parameter is the initial orbital distance. The general trend in the data suggests that minor planets at a greater distance from the star can better retain their water. This is true, as one might naturally expect, for the overwhelming majority of cases. There could be exceptions to this rule, however they require special circumstances. Specifically, these exceptions occur for rock-rich minor planets around 2 M$_{\odot}$ progenitors, as seen in Panel \ref{fig:2sm}. The rightmost subplot shows the water retention for a rock/ice mass ratio of 3. Note the solid blue line that shows the water retention for minor planets with a radius of 25 km and a formation time of 3 Myr. It can be seen that water retention decreases from 175 AU to 20 AU, however it then sharply increases again as the initial orbital distance shortens. The explanation in this case is that at a distance of less than 20 AU, the surface temperature of the minor planet increases. This change in the boundary condition also imposes higher internal temperatures, which in turn allow for some small fraction of the internal rock to hydrate (react with liquid water) during the minor planet's early evolution. Hence, below 20 AU the retention of water actually increases towards the star. This trend is again reversed below 7 AU, since the surface temperature is so high that ice is expelled even prior to attaining hydration temperatures. We thus get peak water retention at 7 AU for this particular combination of parameters. All other minor planets on the same subplot which have a radius larger than 25 km, will also sublimate all their internal ice below a distance of 20 AU. Nevertheless, in these larger objects water retention may behave differently. For example, in a minor planet with a radius of 100 km, water retention below 20 AU actually decreases. The reason in this case is that the peak internal temperatures during the first few Myr of evolution are much higher (due to the larger radius), and may even exceed the point of rock dehydration, in which the rock exudes some of the water it had previously absorbed. The change in surface boundary condition at closer distances again imposes higher internal temperatures, which is why more dehydration occurs towards the star (hence the reduction in water retention).

We are thus faced with the recognition that water retention depends on the particular combination of model parameters. The early evolution of a minor planet determines much of its water retention outcome: its internal structure, whether or not it is differentiated, and how much of its rock is hydrated or dehydrated. During the main sequence and post-main sequence, the luminosity of the star then determines if ice is removed via sublimation, either partially or completely. Note that here we only consider water ice, although other, far less abundant yet more volatile ices may also exist, especially in the smallest minor planets.

Figs. \ref{fig:2-7-3-3} - \ref{fig:6.4-5-5-2} demonstrate exactly how water retention is affected by various combinations of parameters. Fig. \ref{fig:2-7-3-3} shows the compositional cross-section of a minor planet with a radius of 100 km, following its first 10 Myr of evolution around a 2 M$_{\odot}$ star (since this is an animated figure, one may view the evolution, in addition to the end state which is depicted by the still image). In this case the formation time is short and the composition is rock-rich, so the minor planet quickly differentiates. Almost all the anhydrous rock becomes hydrated, embedding nearly half of the initial water content onto it. The remaining water forms a very thin ice-rich crust which later sublimates via insolation from the star. For progenitor star masses of 3 M$_{\odot}$ and above, these stars are so massive that most or even all the ice inside large minor planets close to the star, is sublimated prior to reaching hydration temperatures. Fig. \ref{fig:3-12-4-3} shows the compositional cross-section of a minor planet with a radius of 100 km, following its first 1.7 Myr of evolution around a 3 M$_{\odot}$ star. In this particular example some hydration still occurs, however a considerable fraction of the rock remains anhydrous (in the animation it can be seen that outside-in sublimation of ice from the mantle and inside-out differentiation operate on competing time-scales). In some very extreme cases water can be retained internally even though the minor planet is close to the star, but not as a result of rock hydration. This occurs only for in minor planets with a long formation time and a water-rich composition, around extremely massive stars (5-6.4 M$_{\odot}$, see pink dashed line in Panels \ref{fig:5sm} and \ref{fig:64sm}). In this heating regime the star reaches the WD phase so quickly that despite its intense luminosity some fraction of the internal ice remains. E.g., Fig. \ref{fig:6.4-5-5-2} shows the compositional cross-section of a minor planet with a radius of 100 km, following its entire 64 Myr evolution to the WD stage, around a 6.4 M$_{\odot}$ star. It can be seen that the inner core retains its original icy composition, simply because heat does not have sufficient time to reach the centre. We note that forming a water-rich minor planet at close distances around such massive stars is probably unrealistic, given their short lifetimes and their distant snow lines \citep{KennedyKenyon-2008}. Therefore this combination of parameters is perhaps only plausible for a minor planet that has migrated inward.

Another trend in the data suggests that while more massive stars have a shorter lifetime, and a higher final/initial mass ratio (hence minor planets experience more orbital expansion), their high luminosities clearly dominate the fate of water. We define the outer bound of water retention as the distance beyond which the initial amount of water is fully retained. Examining the smallest minor planet in our sample (radius=1 km), it is immediately evident from these plots that the outer bound of water retention increases steadily with progenitor mass. 

The inner bound of water retention is similarly defined as the minimal distance above which water retention is greater than zero. The behaviour of the inner bound of water retention is a bit more complex compared to the former, as illustrated by Figs. \ref{fig:Masses} and \ref{fig:RetentionVsMass}, however the general trend is also that the inner bound increases with progenitor mass. In 1 M$_{\odot}$ progenitors, the exact fraction of water retained at the inner bound (3 AU) as a result of silicate hydration can be much larger than zero, and it depends on the precise parameters of the minor planet. More massive progenitors than about 2 M$_{\odot}$ already cannot retain any water at 3 AU, which only illustrates the importance of progenitor mass. 

We also investigate the effect of metallicity on water retention. We limit the investigation to 2 M$_{\odot}$ progenitor stars, which approximately correspond to the peak mass in the WD mass distribution (see Section \ref{S:Intro}). In Fig. \ref{fig:Metallicities} we investigate their water retention for [Fe/H]=-1, and [Fe/H]=-2, that is, one order of magnitude and two orders of magnitude reduction in metallicity compared with Panel \ref{fig:2sm} of Fig. \ref{fig:Masses}. Clearly the progenitor metallicity is of lesser importance than the progenitor mass. For example, in Panel \ref{fig:2vlm} the final WD mass is 0.66 M$_{\odot}$, more than but similar to the resulting WD mass in Panel \ref{fig:3sm} (0.65 M$_{\odot}$), and the evolution time of the former is longer than that of the latter. So while an extremely metal-poor progenitor can easily result in a comparable WD mass to a considerably more massive progenitor, it nevertheless does not affect water retention nearly as much, despite having a longer lifetime.

\section{Discussion}\label{S:discussion}
In this study we investigate a wide range of progenitor masses, relevant to G, F, A and B type stars, and also a range of metallicities, from solar abundance down to 10$^{-2}$ solar abundance. The results in Section \ref{S:Results} indicate that the progenitor mass is inversely correlated with water retention, and that water-bearing planetary systems around very massive stars in the range 5-6.4 M$_{\odot}$, if in fact they exist, cannot retain much of their water even to Kuiper Belt distances. Less massive progenitors in the mass range 3-3.6 M$_{\odot}$ result in intermediate water retention in their planetary systems. The innermost minor planets that retain water do not, however, include large quantities of water in the form of hydrated silicates (at least up to the size investigated in our sample). The least massive, yet most common WD progenitors in the mass range 1-2 M$_{\odot}$, allow for the highest degree of water retention in their respective systems, including plenty of hydrated silicates in the interiors of large minor planets at all distances. This is particularly true for 1 M$_{\odot}$ that can retain hydrated silicates even down to a minimal initial orbital distance of 3 AU.

\begin{figure*}
	\begin{center}
		\subfigure[Minor planet radius = 1 km]
		{\label{fig:fin1}\includegraphics[scale=0.5]{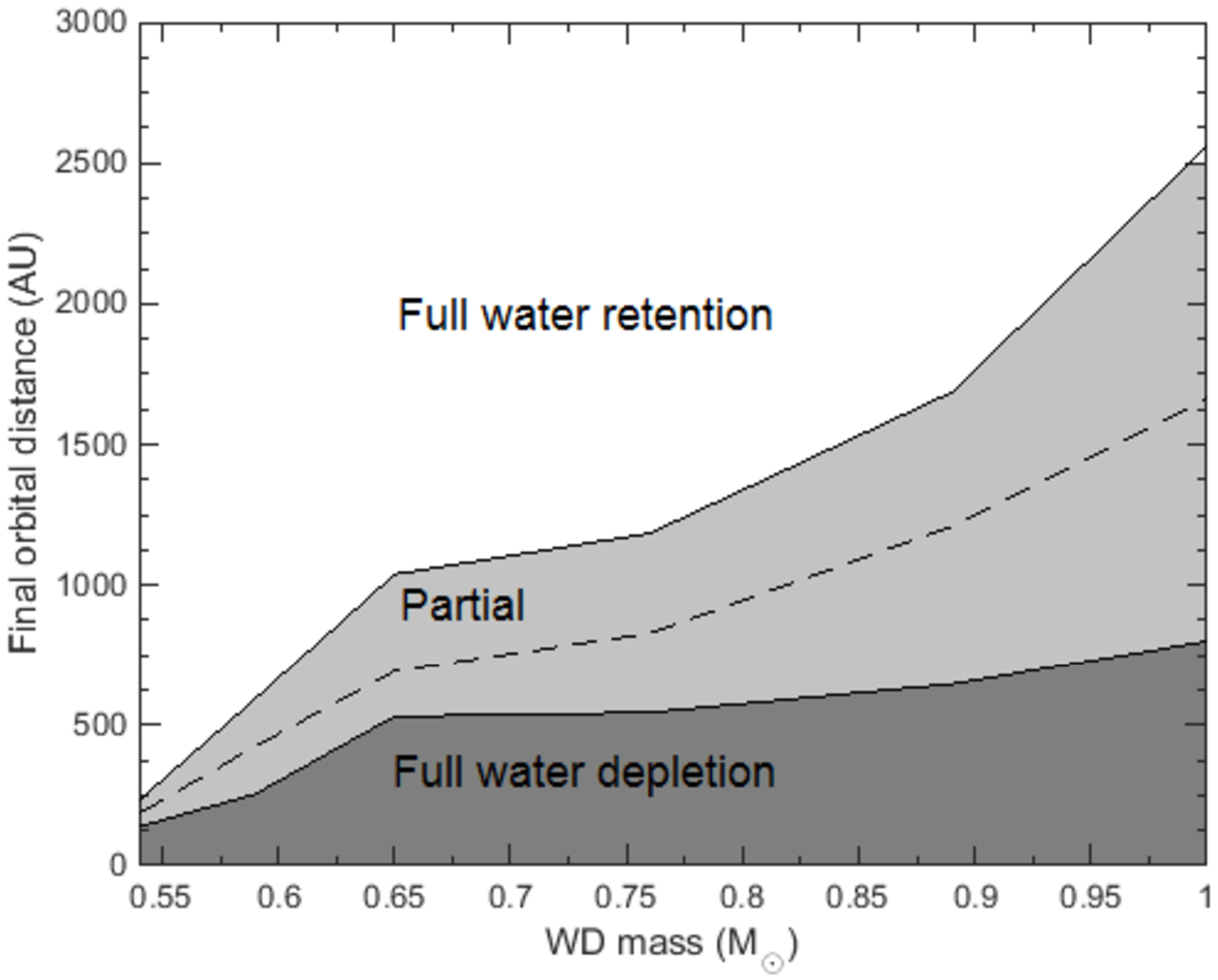}}
		\subfigure[Minor planet radius = 5 km]
		{\label{fig:fin5}\includegraphics[scale=0.5]{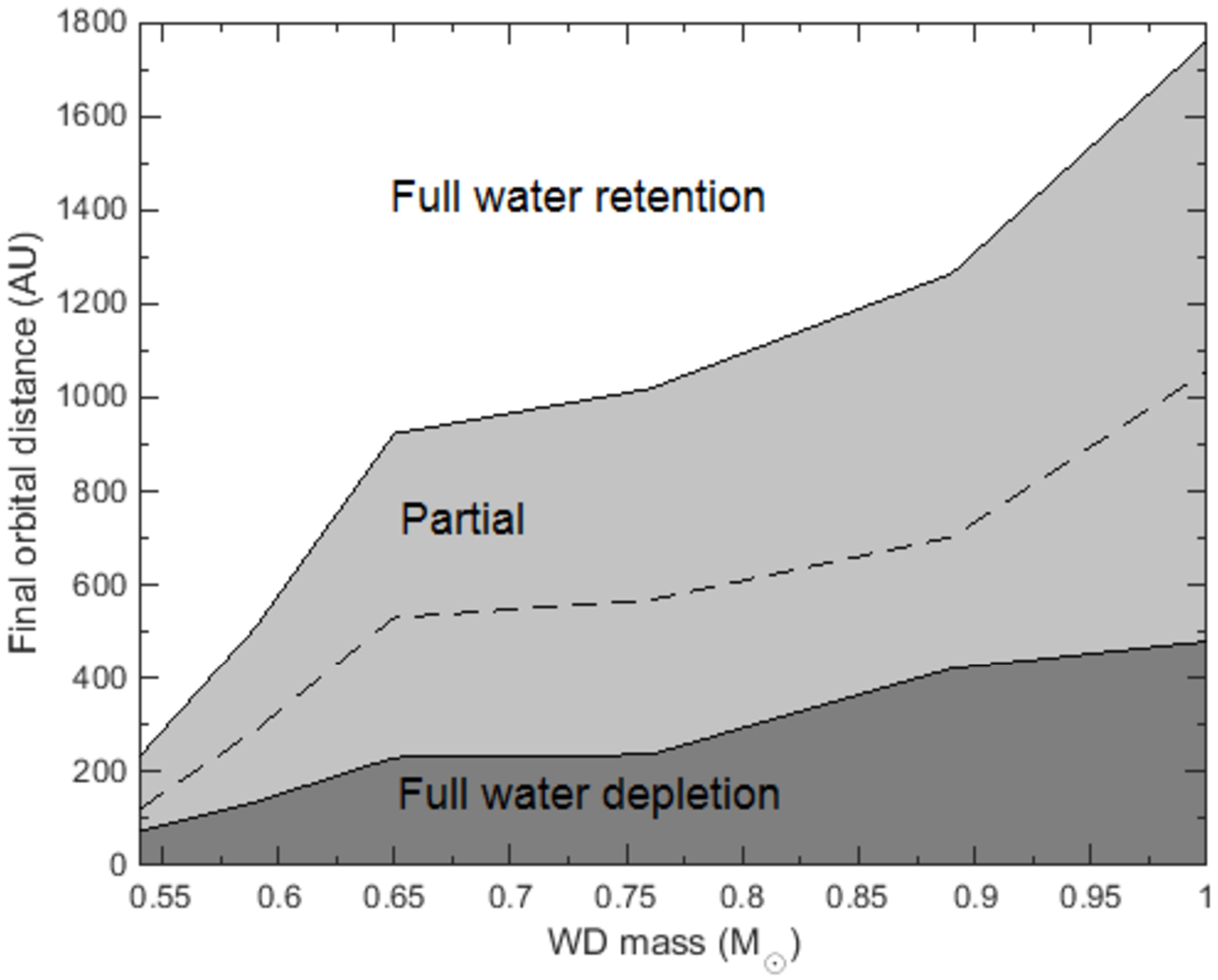}}
		\subfigure[Minor planet radius = 25 km]
		{\label{fig:fin25}\includegraphics[scale=0.5]{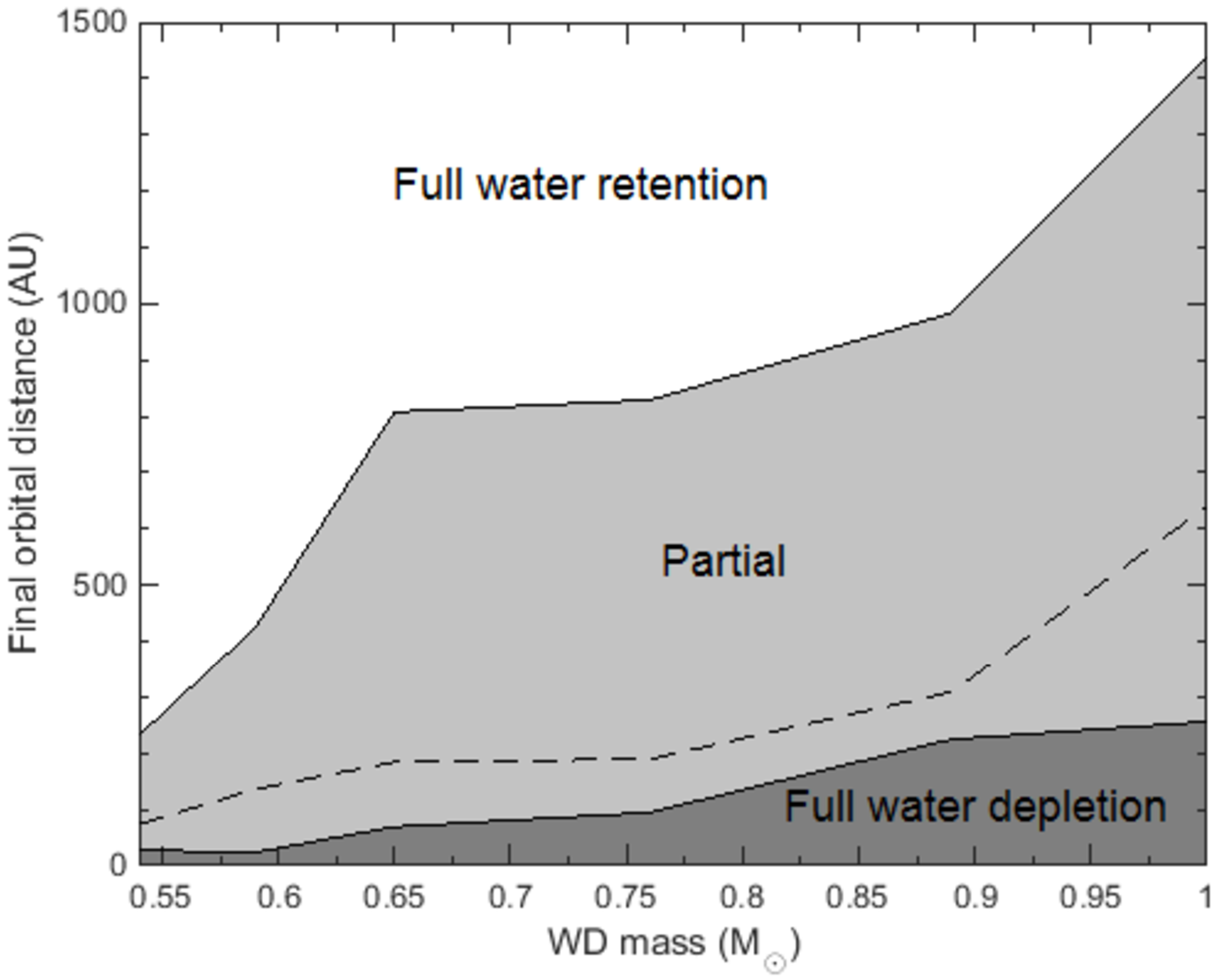}}
		\subfigure[Minor planet radius = 50 km]
		{\label{fig:fin50}\includegraphics[scale=0.5]{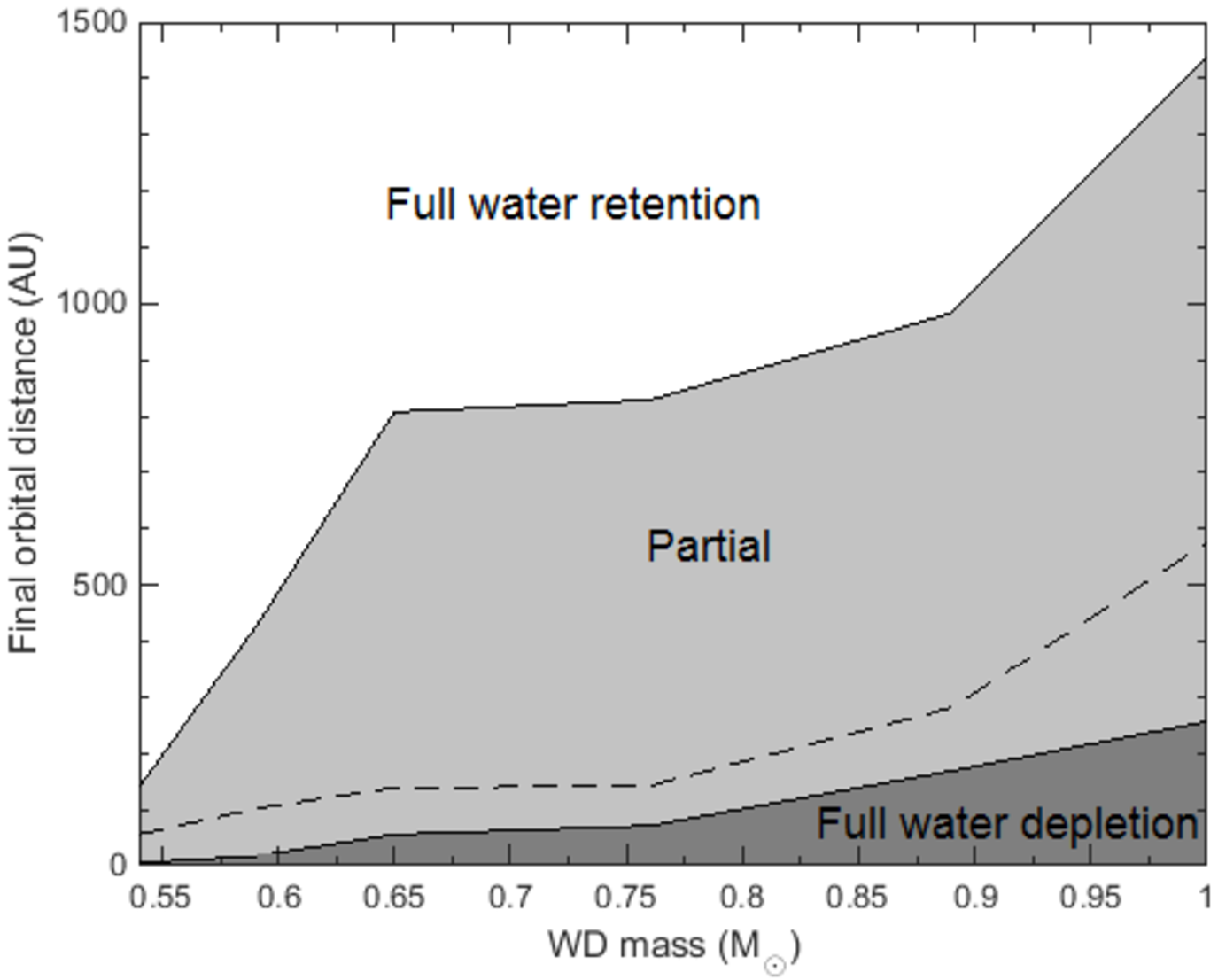}}
		\subfigure[Minor planet radius = 100 km]
		{\label{fig:fin100}\includegraphics[scale=0.5]{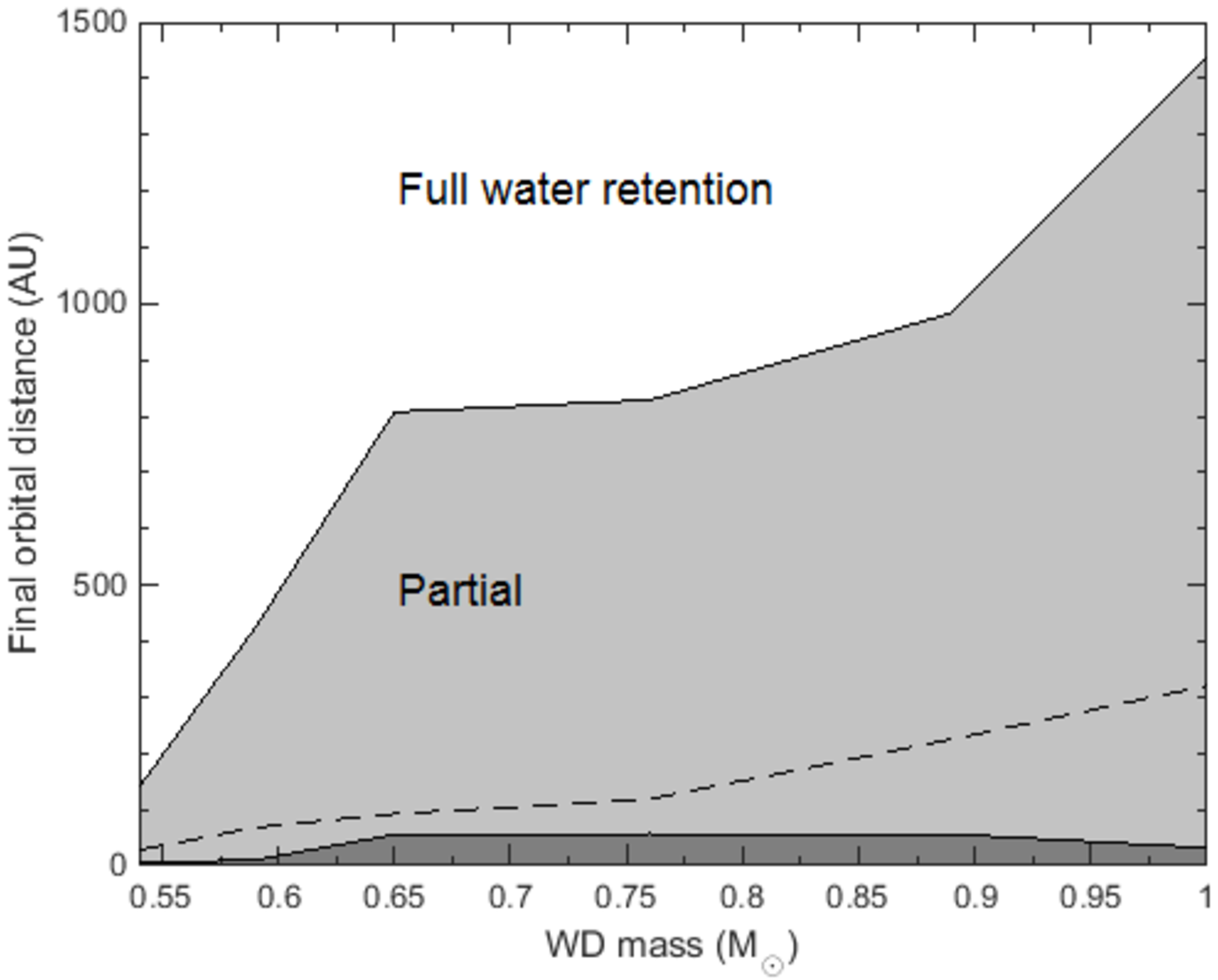}}
	\end{center}
	\caption{Water retention as a function of WD mass. The top and bottom solid lines in each panel mark the outer and inner water retention bounds respectively, expressed in terms of the minor planet's final orbital distance. Above the top solid line minor planets retain their original water content, and below the bottom solid line no water survives. The grey area in between the solid lines represents partial water retention, where 50$\%$ water retention is marked by the dashed line.}
	\label{fig:RetentionVsMass}
\end{figure*}

Fig. \ref{fig:RetentionVsMass} shows the dependence of water retention (in terms of the final orbital distance of a minor planet, that is, after orbital expansion) on the WD mass. Panels \ref{fig:fin1}-\ref{fig:fin100} refer to minor planets of increasing size. Note that here we consider approximate mean values, averaging over different formation times and compositions, given same-size minor planets. Our analysis of different progenitor masses clearly indicates that there is a water retention outer bound, a distance beyond which water is fully retained. Similarly, the water retention inner bound is the minimal distance above which water retention is greater than zero. Here we find both the outer and inner bounds (top and bottom solid lines respectively) to generally increase with WD mass. The only exception is for the inner bound in Panel \ref{fig:fin100} which is shown to decrease for very massive WDs, however this is only true for the \emph{mean} value, whereas we recall that icy minor planets with short distance orbits and long formation times are not very likely.
	
Remarkably, in very massive WDs, one expects to find comets with full or even partial water retention only at distances of thousands, or hundreds of AU, respectively. This find has implications for WD pollution, since any delivery mechanism for these exo-comets onto the WD \citep{VerasEtAl-2014b,StoneEtAl-2015,CaiazzoHeyl-2017} has to be compatible with such great distances. Much larger minor planets (radius $>$ 25 km), primarily around WDs that are 0.6 M$_{\odot}$ or less, can retain a large fraction of their water at relatively close distances (e.g., the 50$\%$ retention dashed line is approximately less than $\sim$100 AU given these conditions), which could also be significant in constraining their delivery mechanism \citep{DebesEtAl-2012,BonsorEtAl-2011,FrewenHansen-2014,DebesSigurdsson-2002,MustillEtAl-2014,VerasGansicke-2015,VerasEtAl-2016,PayneEtAl-2016,PetrovichMunoz-2016,KratterPerets-2012,PeretsKratter-2012,ShapeeThompson-2013,MichaelyPerets-2014,HamersPortegiesZwart-2016}. 

Given these results, we predict a marked difference in the water pollution of WDs of various masses. Minor planets around less massive WDs should be able to retain more water, and specifically at closer distances to the WD. If everything else is equal, both of these results imply that low-mass progenitor stars should have a higher rate of water-bearing minor planets perturbed onto the WD. This effect might be detectable in the future. For example, note Fig. 5 in \cite{RaddiEtAl-2015} which plots the total mass of hydrogen in the convection zones of helium-dominated white dwarfs as function of T$_{eff}$, or cooling age. One of the possible interpretations for the increase in trace hydrogen with cooling age is the long-term accretion of water-bearing minor planets that contribute hydrogen which does not diffuse out of the WD atmosphere. If this interpretation is correct, then in the future (as more data points become available) a richer statistics might enable us to identify different tracks along this plot, for various WD masses. I.e., we would expect a steeper slope for tracks that bin low-mass WDs.

Finally we show that compared to the variations in progenitor mass, even a two order of magnitude reduction in stellar metallicity results in much smaller differences in water retention. We conclude that future studies should be less concerned with the direct effect of metallicity on water retention.

\section{Acknowledgment}\label{S:Acknowledgment}
We wish to thank the anonymous reviewer for helpful comments. UM and HBP acknowledge support from the Marie Curie FP7 career integration grant "GRAND", the Research Cooperation Lower Saxony-Israel Niedersachsisches Vorab fund, the Minerva center for life under extreme planetary conditions and the ISF I-CORE grant 1829/12.

\newpage


\bibliographystyle{apj} 
\bibliography{bibfile}  

\end{document}